\documentclass[a4paper,fleqn,usenatbib]{mnras}
\usepackage{newtxtext,newtxmath}
\usepackage[T1]{fontenc}
\usepackage{ae,aecompl}
\usepackage{pdflscape}
\usepackage{epstopdf}
\usepackage{graphicx}	% Including figure files
\usepackage{amsmath}	% Advanced maths commands
\usepackage{amssymb}	% Extra maths symbols

\newcommand{\kms}{\,km\,s$^{-1}$}
\newcommand{\msun}{\,M$_{\odot}${}} % Solar masses.
\newcommand{\zsun}{\,Z$_{\odot}$} % Solar masses.
\newcommand{\fesixty}{$\rm ^{60}$Fe}
\def\gr{$\gamma$-ray}
\title[]{Near-Earth supernova activity during the past 35 Myr}
\author[M. S\o rensen et al.]{
Mads S\o rensen,$^{1}$\thanks{E-mail: mads.sorensen@unige.ch (MS)}
Henrik Svensmark,$^{2}$ and
Uffe Gr\aa e J\o rgensen$^{3}$
\\
$^{1}$Observatoire de Geneve, University of Geneva, Chemin des Maillettes 51, Versoix 1200, Switzerland\\
$^{2}$DTU Space, Danmarks Tekniske Universitet, Elektrovej bygning 327, Kgs. Lyngby 2800, Denmark\\
$^{3}$StarPlan, Niels Bohr Institute, University of Copenhagen, \O ster Voldgade 5-7, Copenhagen, 1350, Denmark
}

\date{Accepted XXX. Received YYY; in original form ZZZ}

\pubyear{2017}

\begin{document}
\label{firstpage}
\pagerange{\pageref{firstpage}--\pageref{lastpage}}
\maketitle

\begin{abstract}

%The history of near-Earth supernovae (SNe) in the solar neighbourhood out to a distance of 1kpc has been poorly constrained.
Here we combine observations of open clusters (OCs) with single- and binary population synthesis models and a Galactic potential to reconstruct the SN activity of these OCs during the past 35 Myr.
We find that several OCs potentially hosting SN progenitors have passed within 100 pc of the Sun during the past 35 Myr. In particular we find that ASCC 19, NGC 1981, and NGC 1976 are likely to have hosted one or more SNe while passing within 200 pc of the solar system in the period 17 - 12 Myr BP which might have affected Earths' geology and climate.
Besides the stellar history of the individual OCs we also compute 1) a spatial and temporal 2D-probability density map showing the most likely position and time of SN from our sample of OCs within 1 kpc during the past 35 Myr, 2) the time series of the SN rate per volume and 3) the relative SN rate compared with today and corrected for OC evaporation of older generations.
The SN rate today from core collapse is estimated to be 37.8$\pm$6.1\,$\rm kpc^{-3} Myr^{-1}$. During the past 35 Myr we find a peak SN rate around 10 Myr before present (BP) where the rate was 40\% higher relative to the past 1 Myr. Finally we discuss possible effects of binary stellar evolution in relation to the history of SN production in the solar neighbourhood and the detected \fesixty{} signal in terrestrial geological samples induced between $\sim$2.2 - 2.8Myr BP.
\end{abstract}

\begin{keywords}
Stellar evolution -- Open Clusters -- Milky Way -- Binary stars -- Supernova - astrobiology -- 
\end{keywords}

%%%%%%%%%%%%%%%%%%%%%%%%%%%%%%%%%%%%%%%%%%%%%%%%%%

%%%%%%%%%%%%%%%%% BODY OF PAPER %%%%%%%%%%%%%%%%%%

\section{Introduction}
A near-Earth supernova (SN) may influence climate and life on Earth by bombarding the Earth's atmosphere with \gr s and an increased influx of cosmic-ray particles \citep{Terry:1968gc,Ruderman:1974gy,whitten76,ellis93,gehlers03,svensmark12,thomas16}.
Though potentially lethal, near-Earth SNe are rare. To see this, we first define the SN frequency as the number of SNe within the Galaxy pr. century. Secondly, the SN rate is defined as the volume limited number of SN per time in units of $\rm kpc^{-3} Myr^{-1}$. Thirdly, we define a near-Earth SN as any SN within 1 kpc of the Sun.

The mean time between two near-Earth SNe, using the Galactic SN frequency of $2.8\pm0.6$ SNe per 100~yr \citep{li2011}, can be estimated as follows. We begin by assuming an exponential stellar disc with a half-length $H=4$kpc, truncated at an outer distance of $R_{\rm out}=$20 kpc. With the Sun being positioned at R=8.5 kpc from the Galactic centre, the expected time between SNe within 1 kpc of the Sun is 0.02 Myr, and within 100 pc is $\sim$1.8Myr. A SN within 10 pc is only expected to occur ones every $\sim$183 Myr. The simple estimate for near-Earth SNe given above assumes a axis-symmetric, static, and radial exponential distribution of SN progenitors within the Galactic disk which on very long time scales might be reasonable. However, on a stellar evolution time scale of massive stars (10-40 Myr) local fluctuations are expected. The fluctuations occur because stars are born within embedded clusters to form OB associations and OCs which house a significant fraction of all Galactic SN progenitors. This clustering of massive stars means that locally the SN rate can be much higher than that expected from the Galactic SN frequency. Hence, when the Sun passes through such active star forming regions the number of near-Earth SNe can increase dramatically.

In order to estimate the near-Earth SNe in the past one must account for such local fluctuations in the SN production. On short time scales of a few Myr one can look at the distribution of local OB associations. But the short lifetime of O stars and OB associations in general, limits the study based on OB associations to $\approx$5 Myr. Such studies show that the closest region of on-going star formation is located in the Sco-Cen OB association somewhat beyond the 100~pc distance from the Earth. This region has been conjectured \citep{benites02,breitschwerdt16} to be the most probable location of the recent near-Earth events, responsible for the deposition of $^{60}$Fe isotope on the Earth and the Moon \citep{knie99,knie04,Fitoussi:2008cb,wallner16}. In support of one or more recent SN related to the Sco-Cen OB association are the back tracing and pairing of radio quiet pulsars observed in X-ray with their parent cluster, which suggests that their time of ejection from the Sco-Cen OB association was when this group of stars was within 100 pc of the Sun \citep{tetzlaff2010}. However, another source such as the OC Tuc-Hor is an alternative candidate to have produced the recent near-Earth SN \citep{mamajek16,fry16}.

A second probe of recent and historical star formation is OCs, which are gravitational bound long lived stellar groups. OCs have a mean life time of a few hundred Myr \citep{galdyn2008}. The number density of OCs in the solar neighbourhood as a function of their stellar age reliably traces the star formation history up to 500 Myr BP \citep{delaFuenteMarcos:2004jo,svensmark12}. The caveat of using this distribution is the relative low resolution (8-10Myr) of reproduced history which only captures the grand scale star formation history such as the solar neighbourhood passing a spiral arm. Statistically one expects an increased probability of near-Earth SNe following spiral arm passages. However, a near-Earth SN due to a specific OC coming close to the Sun while housing one or more SN progenitors is not caught. To better capture potential past near-Earth SNe from a specific OC it is necessary to know its past position relative to the Sun and also estimate the most likely evolution of its stellar content.

Of all stars born only about one in a thousand is expected to be a core collapse SN progenitor. It is generally believed that the critical zero age main sequence (ZAMS) mass of a SN progenitor is 8-9\msun{} and lives for $\sim$40 Myr before core collapse. Indeed this is true for single star evolution where the lifetime of the star is inversely proportional to its mass cubed. This means, that in a star burst no core collapse SN can happen beyond a few tens of Myrs. However, if one includes binary evolution effects such as mass transfer and stellar mergers a core collapse SN can happen much beyond the 40 Myr. As was recently shown by \citet{zapartas2017} the delay time function of core collapse SN progenitors including binary evolution is 200-250 Myr and the minimum mass at ZAMS of a SN progenitor drops to $\sim$4\msun. \citet{zapartas2017} find that statistically $\sim$85\% of core collapse SNe are still happening during the first $\sim$40Myr with the remaining fraction taking place between 40 and 250 Myr. A special type of SN are Type Ia the mechanism of which is largely unknown though it is believed that it is not a core collapse SN but something else like ignition of a white dwarf due to mass transfer from a close companion or the merging of two white dwarfs. \citet{cappellaro1997} and \citet{smartt2009} estimate SN of type Ia is estimated to be 0.24-0.27 respectively, hence less common than core collapse SN. In the present study we will neglect SN of type Ia since these are generally associated with much older populations of stars and through out the remainder of this paper we mean core collapse SN when we write SN.

Here we wish to constrain the past SN history of the solar neighbourhood using observationally inferred data of OCs. We do this by applying a Galactic gravitational potential with single- and binary star population synthesis modelling to perform a Monte Carlo (MC) study of the near-Earth SN from OCs during the past 35 Myr. We use the age and current mass of OCs as input to a semi-analytic model of OCs mass decay to estimate their birth masses. The stellar population in each OC is reconstructed from up-to-date single- and initial binary property distribution functions from \citet{moe2016}. Hereby we explore the solar neighbourhood SN history some 35~Myr back in time out to a distance of $\sim$1kpc.

Our paper is build up as follow. In section 2 we introduce our compiled data. Section 3 describes our model setup and how these are linked to estimate past SN events near the Sun. In Section 4 we present our overall results and discuss noteworthy OCs. Sections 5 and 6 are devoted to discussion of our results and our conclusion respectively.

\section{Data}
From the catalogue compilation of \citet{Wu:2009} we obtain current sky positions, distances, proper motions, and radial velocities. This catalogue contains a total of 488 OCs. Ages and masses of each OC are taken from the catalogue of \citet{Piskunov:2007bg} which contains a total of 650 OCs. Combining the two catalogues gives us a sample of 395 OCs for which we have a complete set of present-day conditions required to reconstruct the past SN events. Since there is no uncertainty listed in the distance to each OC we set this to be $\pm20\,\%$ of the distance as was also done by \citet{Wu:2009}. In table \ref{tab:data}, a subset of our compiled data set is shown.
% table with most prominent OCs and their yields.
\begin{table*}
	\centering
	\caption{Subset of our sample of 395 OCs. From left to right the columns shows; OC name, age in Myr, logarithm of the mass in units of \msun, sky coordinates $\alpha$ and $\delta$ in degrees, distance $D$ from the Sun in kpc, proper motions $\mu_{\alpha}\cos(\delta)$ and $\mu_{\delta}$ in units of mas yr$^{-1}$, and the radial velocity in \kms.}
	\label{tab:data}
	\begin{tabular}{lccccccccc} % four columns, alignment for each
		\hline
		& Name & $\tau$ & $\log M(\tau) $ & $\alpha$       & $\delta$       & $D$  & $\mu_{\alpha}\cos(\delta)$ & $\mu_{\delta}$     & $V_{r}$ \\
        &     & Myr    &  \msun{} &$\rm ^{\circ}$ & $\rm ^{\circ}$ & kpc & mas\,$\rm yr^{-1}$         & mas\,$\rm yr^{-1}$ & km\,$\rm s^{-1}$ \\
		\hline
		1 & ASCC 1 & 178$\pm$ 36&4.0$\pm$0.2 &  2.40 & 62.67 & 4.00$\pm$0.80 & -2.07$\pm$0.72 & 0.46$\pm$0.57 & -69.70$\pm$4.70 \\
2 & ASCC 3 &  79$\pm$ 16&2.8$\pm$0.4 &  7.79 & 55.27 & 1.70$\pm$0.34 & -1.92$\pm$0.61 & -1.25$\pm$0.59 & -37.00$\pm$0.00 \\
3 & Alessi 20 & 166$\pm$ 33&1.4$\pm$0.2 &  2.35 & 58.65 & 0.45$\pm$0.09 & 8.73$\pm$0.53 & -3.11$\pm$0.53 & -11.50$\pm$0.00 \\
4 & Blanco 1 & 209$\pm$ 42&3.5$\pm$0.2 &  1.03 & -28.17 & 0.27$\pm$0.05 & 20.17$\pm$0.51 & 3.00$\pm$0.51 & 4.10$\pm$1.40 \\
5 & IC 1590 &   7$\pm$  1&1.7$\pm$0.4 & 13.20 & 56.62 & 2.94$\pm$0.59 & -1.36$\pm$0.23 & -1.34$\pm$0.83 & -32.50$\pm$6.40 \\
6 & Mamajek 1 &   8$\pm$  2&-0.4$\pm$0.4 & 130.53 & -78.98 & 0.10$\pm$0.02 & -30.00$\pm$0.30 & 27.80$\pm$0.30 & 16.10$\pm$0.50 \\
7 & Mayer 1 &  55$\pm$ 11&3.1$\pm$0.4 &  5.47 & 61.75 & 1.43$\pm$0.29 & -4.46$\pm$1.13 & -6.66$\pm$0.94 & -20.90$\pm$2.00 \\
8 & Melotte 20 &  35$\pm$  7&3.9$\pm$0.1 & 51.08 & 49.85 & 0.18$\pm$0.04 & 22.41$\pm$0.21 & -25.87$\pm$0.25 & -2.00$\pm$0.10 \\
9 & NGC 129 &  74$\pm$ 15&3.0$\pm$0.2 &  7.50 & 60.22 & 1.62$\pm$0.33 & -1.06$\pm$0.94 & 1.60$\pm$0.94 & -39.40$\pm$0.50 \\
10 & NGC 1981 &  32$\pm$  6&1.3$\pm$0.3 & 83.79 & -3.58 & 0.40$\pm$0.08 & -0.58$\pm$0.69 & 0.80$\pm$0.69 & 27.90$\pm$2.40 \\
$\vdots$ & $\vdots$ & $\vdots$ & $\vdots$ & $\vdots$ & $\vdots$ & &$\vdots$ & $\vdots$ & $\vdots$ \\

		\hline
	\end{tabular}
\end{table*}
The position, distance, proper motion, and radial velocity shown in table \ref{tab:data} are transformed by the method of \citet{Johnson:1987ji} to get heliocentric position and velocity relative to the Sun and this is shown in table \ref{tab:data_transformed}. % table with most prominent OCs and their yields.
\begin{table*}
	\centering
	\caption{Same subset of OCs as in Tab. \ref{tab:data} showing their heliocentric position (kpc) and local standard of rest velocities (\kms).}
	\label{tab:data_transformed}
	\begin{tabular}{lcccccccc} % four columns, alignment for each
		\hline
		& Name & $x$ & $y$  & $z$ & & $u$ &   $v$  & $w$ \\
		&     &   & kpc &   & &   & \kms & \\
		\hline
		1 & ASCC 1 & -1.89$\pm$0.38&3.53$\pm$0.71 & 0.01$\pm$0.00 & &65.82$\pm$106.09 & -43.89$\pm$56.73 & 14.73$\pm$49.06 \\
2 & ASCC 3 & -0.84$\pm$0.17&1.46$\pm$0.29 & -0.22$\pm$0.04 & &32.98$\pm$9.47 & -24.64$\pm$4.83 & -3.92$\pm$6.91 \\
3 & Alessi 20 & -0.21$\pm$0.04&0.40$\pm$0.08 & -0.03$\pm$0.01 & &-9.76$\pm$1.17 & -18.72$\pm$0.63 & -8.80$\pm$1.19 \\
4 & Blanco 1 & 0.05$\pm$0.01&0.01$\pm$0.00 & -0.26$\pm$0.05 & &-23.44$\pm$0.77 & -8.06$\pm$0.66 & -8.87$\pm$1.38 \\
5 & IC 1590 & -1.60$\pm$0.32&2.45$\pm$0.49 & -0.32$\pm$0.06 & &34.56$\pm$29.56 & -18.47$\pm$15.91 & -15.12$\pm$34.26 \\
6 & Mamajek 1 & 0.03$\pm$0.01&-0.08$\pm$0.02 & -0.04$\pm$0.01 & &-11.75$\pm$0.22 & -19.09$\pm$0.44 & -10.52$\pm$0.22 \\
7 & Mayer 1 & -0.70$\pm$0.14&1.24$\pm$0.25 & -0.02$\pm$0.00 & &41.23$\pm$14.33 & -1.49$\pm$7.99 & -41.02$\pm$18.07 \\
8 & Melotte 20 & -0.15$\pm$0.03&0.10$\pm$0.02 & -0.02$\pm$0.00 & &-13.59$\pm$0.17 & -25.69$\pm$0.24 & -7.77$\pm$0.22 \\
9 & NGC 129 & -0.82$\pm$0.16&1.40$\pm$0.28 & -0.07$\pm$0.01 & &25.69$\pm$6.98 & -29.91$\pm$4.27 & 14.70$\pm$9.95 \\
10 & NGC 1981 & -0.33$\pm$0.07&-0.18$\pm$0.04 & -0.13$\pm$0.03 & &-24.07$\pm$2.13 & -10.74$\pm$1.59 & -9.34$\pm$1.46 \\
$\vdots$ & $\vdots$ & $\vdots$ & $\vdots$ & $\vdots$ & &$\vdots$ & $\vdots$ & $\vdots$ \\

		\hline
	\end{tabular}
\end{table*}

\begin{figure}
    \includegraphics[width=\linewidth]{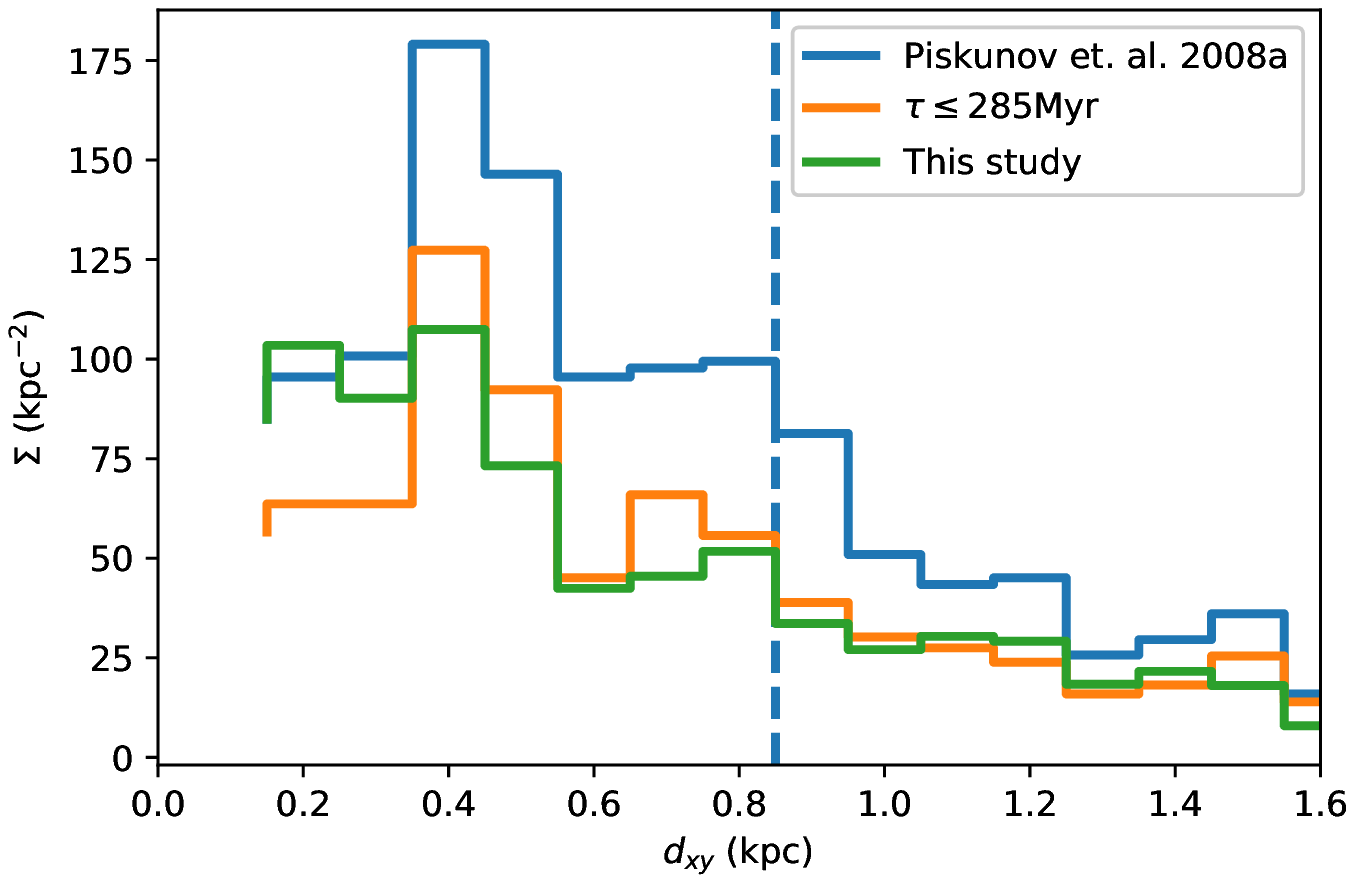}
    \caption{The surface density of OCs with distance to the Sun projected into the $xy$-plane. The blue curve shows the surface density from \citet{Piskunov:2008hp}, the orange curve is similar to the blue curve but only considers OCs with ages $\leq$ 285 Myr, while green curve is the data compiled for this study. The vertical dashed line is at a distance of 850 pc and marks the completeness distance of the catalogue from \citet{Piskunov:2008hp} as was found by \citet{Piskunov:2006gy}. The difference between the blue and green curve in the first bin is due to the OC Collinder 359 having a different distance between the data from \citet{Piskunov:2008hp} and \citet{Wu:2009}.}
    \label{fig:OC_surf_dens}
\end{figure}

In Fig. \ref{fig:OC_surf_dens}, we show the surface density of OCs as a function of distance to the Sun, where the distance is projected into the $xy$-plane. The blue curve shows the surface density for OCs from the data of \citet{Piskunov:2008hp} which is similar to the data compilation used in \citet{Piskunov:2006gy}. The latter is complete out to a distance of 0.85 kpc. Since we are interested in the SN activity during the past 35 Myr, we only consider those OCs with ages less than 285 Myr which are represented by the orange curve in Fig. \ref{fig:OC_surf_dens}. The surface density of the compiled data set for this study is the green curve in Fig. \ref{fig:OC_surf_dens}, and shows good agreement with the orange curve, i.e. we are neaerly complete for young OCs, relevant for the most recent SNe activity within 1 kpc.

\section{Models}
Estimating past SN events from OCs requires accounting for 1) the trajectory of each OC and the Sun within the Galaxy, 2) the mass loss experienced by each OC due to stellar evolution and tidal disruption, 3) the initial distribution of stars in each OC, and 4) the stellar evolution within each cluster for single and binary stars. Below we outline each of these components and end this section by describing how these are combined into our model.

\subsection{The motion of OCs and the Sun in the Galaxy}
\begin{figure}
    \begin{tabular}{l}
    \includegraphics[width=\linewidth]{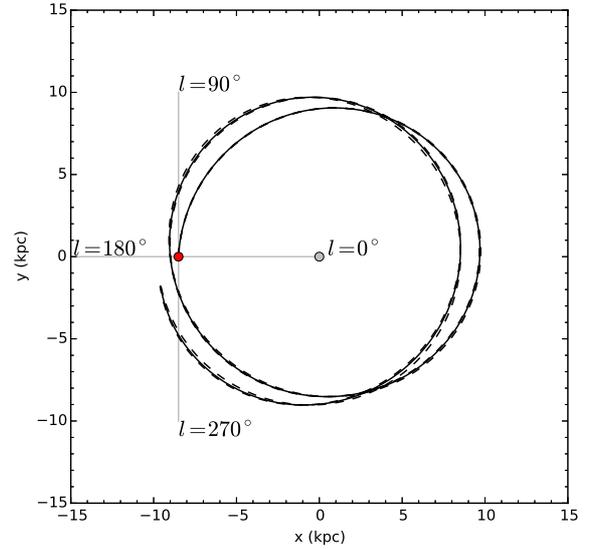}     \\
    \includegraphics[width=\linewidth]{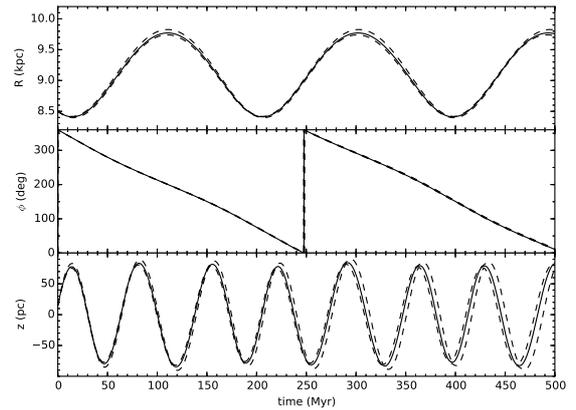}
    \end{tabular}
    \caption{Orbital characteristic of the Sun within the Galaxy for a 500 Myr period. The top panel shows the Suns' position today as a red dot. Drawn on top of the Sun is the Galactic coordinate frame. The bottom panel shows the three cylindrical coordinates of the Sun during its orbit. In both panels, the solid line is the median of 2000 MC simulations and the dashed lines are the 5$^{\rm th}$ and 95$^{\rm th}$ percentiles.}
    \label{fig:solar_motion}
\end{figure}

To trace the trajectories of OCs back in time, we model the Galaxy with respect to a right-handed Cartesian frame of reference in which we solve the equations of motion with an axis-symmetric gravitational potential $\Phi(R,z)$ in cylindrical coordinates $(R,\phi,z)$:
\begin{equation}
\label{eq:EOM}
\begin{split}
\ddot{R} =&  \frac{L_{z}^2}{R^3} -\frac{\partial\Phi(R,z)}{\partial R} \\
\ddot{z} =& -\frac{\partial\Phi(R,z)}{\partial z} \\
L_z      =&  R^2\dot{\phi}
\end{split}
\end{equation}
where $L_{\rm z}$ is the conserved angular momentum. We solve the system of equations in eq. \eqref{eq:EOM} using a $4^{\rm th}$ order Runge-Kutta algorithm \citep{odespy}. We adopt the Galactic potential of \citet{Carlberg:1987fl} with the fit of \citet{Kuijken:1989er}. This potential has the local standard of rest (LSR) position at ($x_{\rm LSR}$, $y_{\rm LSR}$, $z_{\rm LSR}$) = $(x_{\odot}, y_{\odot}, 0.0)$ kpc  which gives it an orbital velocity of 221 \kms.  In the selected frame of reference the Suns' position is $(x_{\odot},y_{\odot},z_{\odot}) = (-8.5, 0.0, 0.015)$\,kpc \citep{Carlberg:1987fl, Kuijken:1989er, Cohen:1995gs}. We adopt the solar motion $(u_{\odot},v_{\odot},w_{\odot}) = (11.1_{-0.75}^{+0.69}, 12.24_{-0.47}^{+0.47}, 7.25_{-0.36}^{+0.37})$ \kms\, with respect to the LSR \citep{Schonrich:2010kt}. The Sun is currently orbiting in the clock wise direction and moving radially inwards while going up and above the Galactic plane. In Fig. \ref{fig:solar_motion}, the motion of the Sun forward in time for a 500 Myr period is shown, based on a MC ensemble of 2000 simulations with 5$^{th}$ and 95$^{th}$ percentile shown. We see that the uncertainty of the Sun's LSR velocity has little effect on the Suns orbit.

\subsection{The initial mass of open clusters}
OCs are groupings of stars bound together by their mutual gravity. Due to effects of stellar evolution, tidal perturbations and kinetic heating, OCs loose their mass via gas loss and ejection of stars. Eventually, OCs become unbound and are dissolved. The time from formation to dissolution of an OC is strongly dependent on its initial mass. Likewise, the current age and mass of an OC is an indication of its initial mass at birth which is important for the initial distribution of stars in that cluster. To estimate the birth mass, $M_{\rm birth}$, of an OC we use the analytic description of \citet{Lamers:2005ip}:
\begin{equation}
M_{birth} = \left\{\left(\frac{M}{M_{\odot}}\right)^\gamma + \frac{\gamma t}{t_0}\right\}^{1/\gamma}\mu_{ev}(t)^{-1}
\end{equation}
with $\gamma=0.62$ and $\tau_{\rm 0} = 2$ Myr. The function $\mu_{ev}(t)$ is defined as $\mu_{ev}(t) = 1- q_{ev}(t)$ where
\begin{equation}
\log q_{ev}(t) = 
\begin{cases}
    0 & \text{if } t \leq 12.5 Myr\\
(\log t - a_{ev})^{b_{ev}} + c_{ev} & \text{if} t > 12.5 Myr \\
\end{cases}
\end{equation}
and describes the fractional mass loss due to stellar evolution with age. For solar metallicity $a_{ev} = 7.0$, $b_{ev} = 0.255$, and $c_{ev} = -1.805$. This description accounts for both mass loss due to stellar winds and evolution, as well as tidal disruption of the clusters.

\subsection{Priors of stellar populations}\label{sec:priors}
\begin{figure*}
	\includegraphics[width=\linewidth]{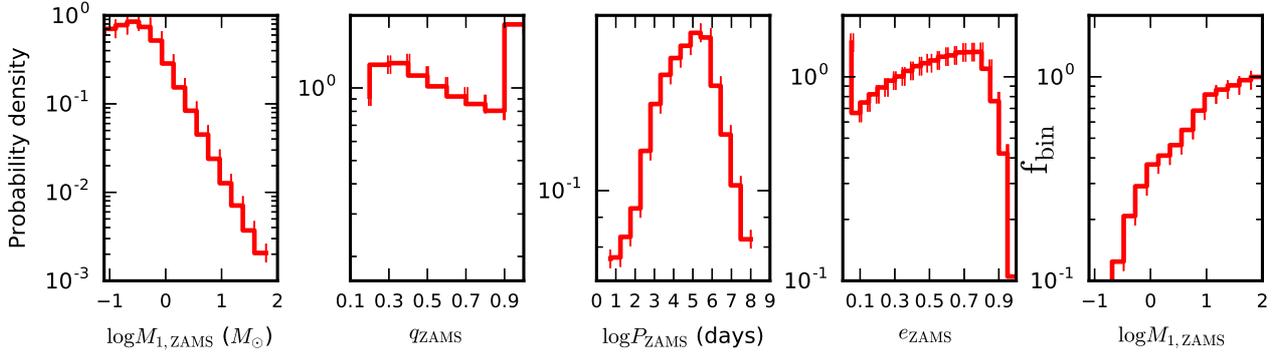}
    \caption{Distributions of initial single and binary properties from the entangled prior distribution \citep{moe2016} from a $10^6$\msun{} star burst. From left to right panel we show the initial stellar mass function, the mass ratio- or $q$-distribution, the orbital period distribution in $\log P$, the eccentricity- or $e$-distribution, and the fraction of binary primaries to single stars as a function of primary mass.}
    \label{fig:priors}
\end{figure*}

Once we know the initial mass of each OC we must determine the number of SN progenitors within each OC. For this purpose we use the prior distribution of single and binary stars from \citet{moe2016} which is given as the joint probability distribution $p(M_1, q, logP, e)$. The function entangles all parameters such that it gives a non-trivial distribution of stars which is different from the product of individual prior distributions. It is the most up to date estimate of how binary stars are distributed in primary mass, mass ratios, orbital periods and eccentricities. Further, it constrains the binary fraction as a function of primary mass. The initial stellar primary mass function, which is the mass distribution of single stars and primaries in binaries is 
\begin{equation}
\frac{dN}{dM_{1,ZAMS}} \propto M_{1,ZAMS}^{\alpha}
\end{equation}
where,
\begin{equation}
\alpha =
\begin{cases}
     -0.8 & \text{if } 0.08 \leq M_{1,ZAMS}/M_{\odot} < 0.5 \\
     -1.6 & \text{if } 0.5 \leq M_{1,ZAMS}/M_{\odot} < 1.0 \\
     -2.3 & \text{if } 1.0 \leq M_{1,ZAMS}/M_{\odot} < 100 \\
    \end{cases}
\end{equation}
This is different from the \textit{traditional} initial stellar mass function (IMF), i.e. the standard or canonical IMF of \citet{kroupa2001, weidner2013}, which is the mass distribution function of single-, primary- and secondary stars.
In Fig. \ref{fig:priors} we show the initial stellar distribution of primaries $M_1$, the companion star mass ratio distribution $q$, the orbital period distribution $\log P$, the eccentricity distribution, and binary fraction with primary mass. The caveat in using the entangled prior is its lack of coverage for mass ratios $q$ < 0.1. However, this mass ratio, though important for e.g. low mass X-ray binaries (LMXB), constitute a small mass component of the full stellar mass. Hence we are only interested in the SN events, and not the later evolution of binaries with a low mass companion.
To deduce the binary fraction with primary mass we sampled the initial stellar population in the primary mass range from 0.08 < $M_{\rm 1,ZAMS}/M_{\rm \odot}$ < 100. From this set of simulations we also deduce that 54\% of the stellar mass is locked up in stellar single and binary systems with primary masses of $\geq$ 2\msun{} which we use to speed up our computation in Sect. \ref{sec:MC}.
% Priors

\subsection{Population synthesis modelling}\label{sec:PopSynth}
\begin{figure}
	\includegraphics[width=\linewidth]{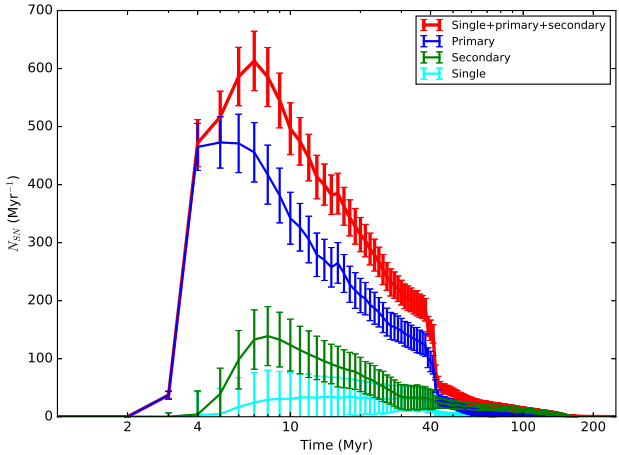}
    \caption{The delay time function or SN response function following a $10^6$\msun{} star burst with metallicity Z =\zsun{}, using the entangled prior introduced in Sect. \ref{sec:priors}. On average the entangled prior produces $1.43\times10^4$ SN during 250 Myr in the the star burst. The uncertainties are the sampling uncertainty at the $5^{th}$ and $95^{th}$ percentile.}
    \label{fig:delay_time}
\end{figure}

To determine the temporal distribution of SNe from each OC we apply a modified version of the Binary Stellar Evolution code \citep[SSE and BSE;][]{hurley2000, hurley2002} that is updated to include the suite of stellar wind prescriptions for massive stars described in \citet{belczynski2010}, the fitting formulae for the binding energy of the envelopes of stars derived by \citet{loveridge2011}, and the prescriptions "STARTRACK", "Delayed", and "Rapid" for compact object (CO) formation in binaries, as described in \citet{fryer2012}. As standard, we will apply a common envelope efficiency parameter, $\alpha_{CE}$ of 0.5 which describes the transformation of orbital energy into kinetic energy of the envelope. For mapping the pre-SN progenitor mass to CO mass we use the Rapid convective SN engine, and a Maxwellian natal kick distribution with max likelihood $\sigma_{v} = 265$\kms for neutron stars (NS). During black hole (BH) formation the potential natal kick is scaled with the amount of fall back mass they may experience during the SN explosion \citep{Belczynski2008}.
Though BSE is a parameterised stellar evolution code, its use is justified for three reasons, first of all it accounts for all relevant phases of single and binary stellar evolution from ZAMS to single star SN explosions and SN in binaries affected by mass transfer and mergers. Secondly, it is computationally efficient for use in MC type simulations and its order of magnitude estimate is sufficient to match the uncertainties of the used data. Third and finally, the alternative would be to ignore the binary evolution, adopt more up-to-date and detailed single star tracks and not being able to estimate the effect of local solar neighbourhood SN rates that include the effects of interacting binaries in SN production.
In Fig. \ref{fig:delay_time} we demonstrate the effect of accounting for the binary evolution on a $10^{6}$\msun{} star burst at Z = \zsun{} by showing the number of SNe explosions produced given the entangled prior distribution. For a discussion on the processes beyond the core-collapse SN of the first 40 Myr phase, such as mergers- or mass transfer induced core collapse SNe, see \citet{zapartas2017}. We note that on average, during a star burst of  $10^6$ \msun, our binary population synthesis produces $1.43\times10^4$ SNe within 250 Myr. The fraction of "delayed" SNe, i.e. $T_{SN}>40$Myr is $14.9\pm0.5\%$ which is in agreement with \citet{zapartas2017}. We have decomposed the distribution into three components; primary, secondary, and single: Primary refers to the most massive star of a binary at ZAMS that go SN. Secondary refers to those binary companions that one way or another goes SN, though we do not keep track of the mechanism that took place. Single refers to those single stars that have ZAMS masses above $8 M_{\rm \odot}$.
% Delay time distribution

\subsection{Doing the Monte Carlo}\label{sec:MC}
We perform MC simulations by combining the trajectories of OCs with evolving single- and binary stellar populations. This allows us to extract the spatial and temporal distribution of SNe. Here we will assume all errors in tables \ref{tab:data} and \ref{tab:data_transformed} are Gaussian.
We feed the initial conditions from Tab. \ref{tab:data_transformed} into the simulation, and for each OC the next steps are the following:
\begin{enumerate}
    \item Randomly vary initial conditions $\vec{v}_{\odot}$,$\vec{r}_i$, $\vec{v}_i$, $\tau_{i}$, $M_{i}$.
    \item Calculate the trajectory of the OC and the Sun back in time
    \item Find the Euclidean distance
    \item Determine initial mass $M_{\rm birth}$
    \item Randomly draw a stellar population
    \item Evolve the stellar population with BSE
    \item Extract all SNe for each OC happening during the past 35 Myr and within 1 kpc.
    \item Infer distances of SNe by interpolating the time of each SN on to the trajectory of the OC
    \item goto i)
\end{enumerate}
We determine the Euclidean distance as
\begin{equation}
D_i(t) = \sqrt{\left(x_i(t)-x_{\odot}(t)\right)^2+\left(y_i(t)-y_{\odot}(t)\right)^2+\left((z_i(t)-z_{\odot}(t)\right)^2}
\end{equation}
When we fill an OC with its initial stellar population, we first withdraw from the OC birth mass, $M_{birth}$, the fraction of mass made up of low mass stars, i.e. $M < 2 M_{\rm \odot}$, and then fill the OC with primary masses in the range from $2 < M_{\rm 1,ZAMS}/M_{\rm \odot} \leq 100$. Mass ratios, $q$, are drawn from 0.1 to 1.
For each OC, subscript $i$, we define a function $f_i(\vec{v}_{\odot}, \vec{r}_{i}, \vec{v}_{i}, \tau_{i},M_{i})$, which depend on an OCs current position, velocity, age, and mass, and the current motion of the Sun. Integrating $f_i$ describes an OCs' distance to the Sun since it was born, its initial stellar content and stellar evolution from birth until present time,
\begin{equation}\label{eq:MC}
\begin{split}
\rho_i(SN_i(D,t)|\vec{v}_{\odot}, \vec{r}_{i}, \vec{v}_{i}, \tau_{i},M_{i}) =& \int f_i(\vec{v}_{\odot}, \vec{r}_{i}, \vec{v}_{i}, \tau_{i},M_{i})\\
 & d^3\vec{v}_{\odot}d^3\vec{r}_{i}d^3\vec{v}_{i}d\tau_{i}dM_{i} \text{ .}\\
\end{split}
\end{equation}
Note that time is implicit expressed in eq. \eqref{eq:MC} through the OC age $\tau_{i}$. Hence we can extract the most likely time and distance of a SN from an OC since it was born. Summation over all $i$ gives the toal number of SNe produced by these OCs or the combined likelihood of a SN going of at some time with some distance to the Sun. For each OC we have an ensemble of $10^4$ simulations.

\section{Results}

\subsection{Orbit of open clusters}
% Distance to open clusters with time for single stars
\begin{figure*}
	\includegraphics[width=\linewidth]{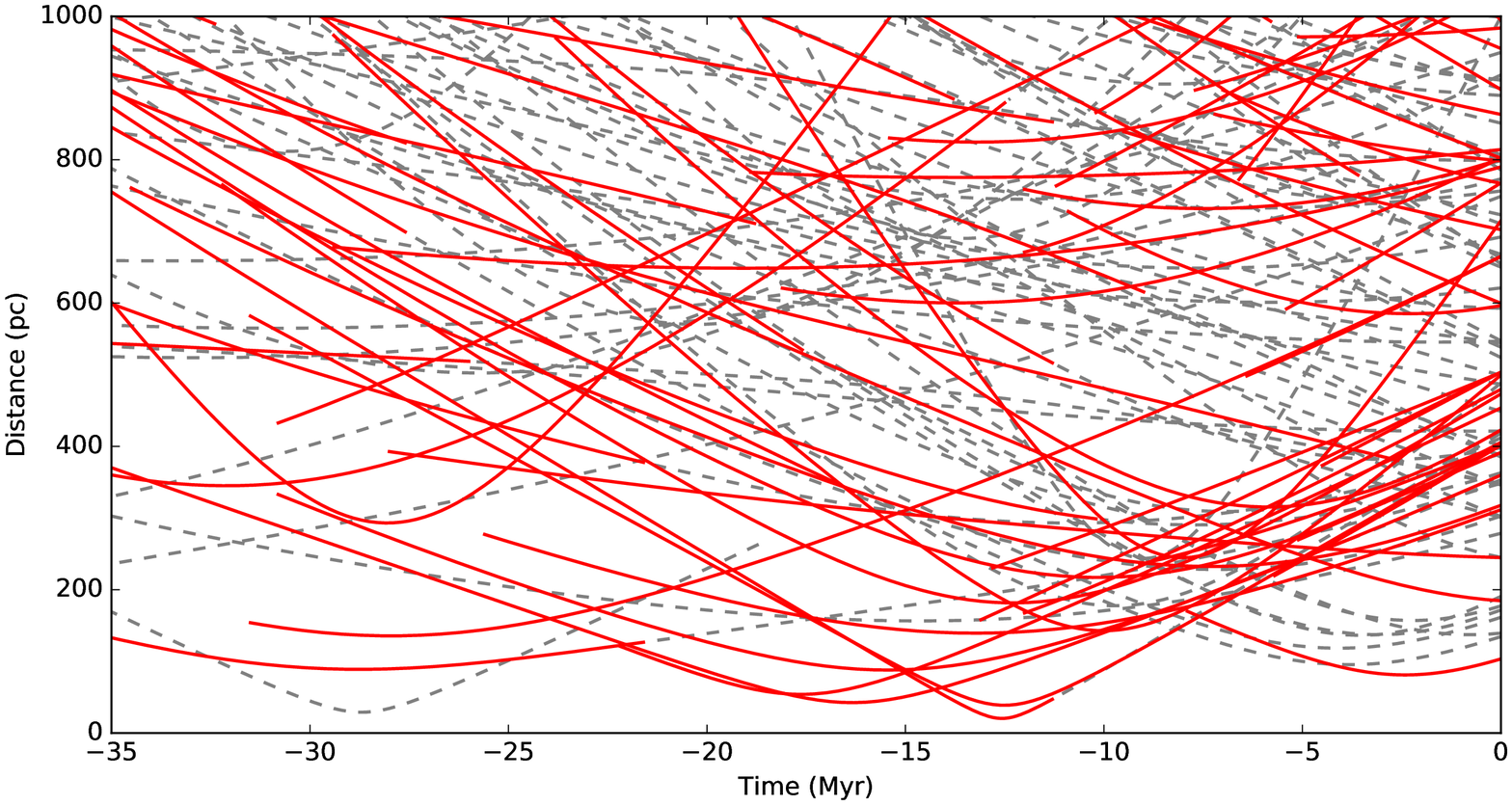}
    \caption{Distance to the Sun of OCs with ages less than 40 Myr (red lines) and OCs with ages 40 < time/Myr $\leq$ 200 (grey dashed lines) for the past 35 Myr, based on nominal values from table \ref{tab:data_transformed}. Binary interactions can extend the time during which SNe occur out to 200-250 Myr. At the earliest times, $\leq$ 40 Myr, SN progenitors come primarily from the evolution of single stars.}
    \label{fig:oc_distance}
\end{figure*}
Figure \ref{fig:oc_distance} shows the distance to the Sun of individual OCs for the past 35 Myr. The distance vs time is decomposed into two time regimes, namely ages < 40Myr (red lines) where SNe explosions is due to core collapse and a regime between 40-200 Myr (grey dashed lines) during which SNe explosions will be affected by binary evolution either by merging or mass transfer. We identify the following interesting features: From -18 to -12 Myr, 5 OCs come within 100 pc of the Sun while potentially housing SN progenitors. Around -10 to -5 Myr there is a grouping of OCs some 200-300 pc from the Sun. This group of OCs is seen moving away from the Sun. A single OC is roughly at a distance of 100 pc between -3 and -2 Myr, while also young enough to house SN progenitors. No OC appears to enter within \textit{the kill radius} of SN explosions of roughly $\approx$10 pc \citep{fry2015}. In table \ref{tab:nearestOC} the overall result of our MC simulation for a subset of all 395 OCs is shown and In sect. \ref{sec:individual_ocs} we discuss 11 of these OCs. The table shows the OCs mean $\pm$1$\sigma$ age, minimum distance $D_{min}$, time and age of minimum distance, and birth mass $M_{\rm birth}$. Also shown are the median number of SN in each OC and $\pm 1\sigma$ percentiles. Finally, we show are the probability of having a single SN within 1000\,pc, 800\,pc, 600\,pc, 400\,pc, 200\,pc, 100\,pc. The lower and upper $1\sigma$ percentile we define as the 16\% and 84\% percentile respectively.

\begin{landscape}
\begin{table}
    \centering
    \begin{tabular}{llcccccccccccc}
\hline
   & Name       &      $\tau$       &   $D_{min}$     &  $t(D_{min})$   & $\tau(D_{min})$ &   $M_{\rm birth}$ &  $N_{SN}$ & P(SN)  & P(SN) & P(SN)& P(SN)& P(SN)& P(SN)  \\
   &            &      Myr       &    pc           &  Myr BP        &     Myr        &  $M_{\rm \odot}$  &  & D<1kpc & D<0.8kpc& D<0.6kpc& D<0.4kpc& D<0.2kpc & D<0.1kpc \\
\hline
1 & Mamajek 1 &  8.1$\pm$ 1.6 & $80.9_{- 9.9}^{+11.2}$ & -2.4$\pm$ 0.4 &  5.7$\pm$ 1.7 &  6.7$\pm$ 2.1 & $1.00_{-1.00}^{+0.00}$ & 0.09 & 0.09 & 0.09 & 0.09 & 0.07 &  -  \\
2 & Melotte 20 & 35.6$\pm$ 7.0 & $185.6_{-25.4}^{+26.4}$ & -0.1$\pm$ 0.3 & 35.5$\pm$ 7.0 & 9105.7$\pm$2404.5 & $87.00_{-24.00}^{+30.00}$ & 0.82 & 0.06 & - & - &  -  &  -  \\
3 & ASCC 19 & 43.6$\pm$ 8.7 & $51.4_{-32.4}^{+81.0}$ & -17.8$\pm$ 8.6 & 25.7$\pm$12.1 & 721.4$\pm$256.9 & $5.00_{-3.00}^{+4.00}$ & 0.95 & 0.80 & 0.61 & 0.17 & 0.01 & - \\
4 & IC 4665 & 42.5$\pm$ 8.5 & $351.2_{-45.6}^{+44.6}$ &  0.0$\pm$ 0.0 & 42.5$\pm$ 8.5 & 872.8$\pm$301.3 & $7.00_{-4.00}^{+4.00}$ & 0.98 & 0.04 & - &  -  &  -  &  -  \\
5 & NGC 2232 & 30.8$\pm$ 6.1 & $103.7_{-33.9}^{+189.8}$ & -12.8$\pm$ 7.9 & 18.0$\pm$ 9.7 & 245.8$\pm$85.2 & $3.00_{-2.00}^{+1.00}$ & 0.92 & 0.80 & 0.63 & 0.18 & 0.02 & - \\
6 & NGC 2547 & 49.9$\pm$ 9.9 & $290.9_{-68.4}^{+67.6}$ & -7.7$\pm$ 0.8 & 42.3$\pm$10.0 & 227.5$\pm$69.3 & $1.00_{-1.00}^{+2.00}$ & 0.69 & 0.27 & 0.08 & - &  -  &  -  \\
7 & Roslund 5 & 58.7$\pm$11.8 & $389.1_{-76.0}^{+74.6}$ & -0.1$\pm$ 1.8 & 58.6$\pm$11.9 & 567.7$\pm$164.3 & $1.00_{-1.00}^{+3.00}$ & 0.50 & 0.06 & - &  -  &  -  &  -  \\
8 & NGC 1981 & 32.0$\pm$ 6.1 & $66.3_{-31.6}^{+36.6}$ & -12.8$\pm$ 3.1 & 19.2$\pm$ 6.8 & 114.7$\pm$48.9 & $2.00_{-1.00}^{+1.00}$ & 0.82 & 0.77 & 0.63 & 0.30 & 0.09 & 0.01 \\
9 & NGC 1976 & 51.2$\pm$10.2 & $57.3_{-29.2}^{+46.0}$ & -12.8$\pm$ 2.9 & 38.4$\pm$10.6 & 352.6$\pm$134.2 & $2.00_{-2.00}^{+2.00}$ & 0.72 & 0.46 & 0.23 & 0.05 & 0.01 & - \\
10 & Trumpler 10 & 24.0$\pm$ 4.7 & $311.6_{-84.4}^{+89.9}$ & -5.5$\pm$ 1.1 & 18.5$\pm$ 4.9 & 722.6$\pm$259.7 & $6.00_{-3.00}^{+4.00}$ & 0.87 & 0.35 & 0.08 & - &  -  &  -  \\
11 & ASCC 20 & 22.3$\pm$ 4.5 & $101.1_{-50.1}^{+284.2}$ & -13.1$\pm$ 7.1 &  9.2$\pm$ 7.9 & 954.9$\pm$369.4 & $7.00_{-3.00}^{+5.00}$ & 0.98 & 0.86 & 0.41 & 0.03 & - & - \\
12 & Platais 8 & 56.2$\pm$11.2 & $94.5_{-22.7}^{+30.5}$ & -3.3$\pm$ 1.0 & 52.9$\pm$11.3 & 1951.4$\pm$446.7 & $6.00_{-5.00}^{+9.00}$ & 0.63 & 0.07 & - &  -  &  -  &  -  \\
13 & Melotte 22 & 120.1$\pm$24.0 & $129.6_{-23.1}^{+24.4}$ & -0.8$\pm$ 0.8 & 119.2$\pm$24.0 & 1504.4$\pm$313.0 & $0.00_{-0.00}^{+0.00}$ & 0.01 &  -  &  -  &  -  &  -  &  -  \\ 
14 & IC 2602 & 67.5$\pm$13.5 & $118.4_{-22.5}^{+28.0}$ & -3.8$\pm$ 1.0 & 63.8$\pm$13.5 & 1334.5$\pm$278.4 & $0.00_{-0.00}^{+5.00}$ & 0.34 & 0.04 & - & - &  -  &  -  \\
15 & Platais 9 & 123.0$\pm$24.6 & $137.7_{-30.3}^{+30.7}$ & -3.1$\pm$ 0.4 & 119.9$\pm$24.7 & 5158.9$\pm$1731.3 & $0.00_{-0.00}^{+0.00}$ & - & - & - & - &  -  &  -  \\
16 & IC 2391 & 76.0$\pm$15.2 & $156.1_{-31.8}^{+34.1}$ & -2.1$\pm$ 0.6 & 73.9$\pm$15.3 & 467.2$\pm$126.0 & $0.00_{-0.00}^{+1.00}$ & 0.03 & - & - & - & - & - \\
17 & NGC 2451A & 57.5$\pm$11.5 & $136.0_{-36.3}^{+36.2}$ & -3.5$\pm$ 0.5 & 54.1$\pm$11.5 & 581.6$\pm$177.4 & $1.00_{-1.00}^{+3.00}$ & 0.23 & 0.02 & - & - &  -  &  - \\
18 & Platais 10 & 158.3$\pm$31.3 & $239.7_{-40.2}^{+38.0}$ & -0.9$\pm$ 2.7 & 157.4$\pm$31.5 & 3774.9$\pm$1298.7 & $0.00_{-0.00}^{+0.00}$ & - & - &  -  &  -  &  -  &  - \\
19 & Platais 4 & 173.8$\pm$34.7 & $154.6_{-90.1}^{+134.4}$ & -7.0$\pm$ 6.8 & 166.7$\pm$35.4 & 2320.6$\pm$702.2 & $0.00_{-0.00}^{+0.00}$ & - & - &  -  &  -  &  -  &  -  \\
20 & NGC 6475 & 166.2$\pm$33.4 & $301.2_{-58.7}^{+59.5}$ &  0.0$\pm$ 0.0 & 166.2$\pm$33.4 & 2329.7$\pm$490.6 & $0.00_{-0.00}^{+0.00}$ & - &  -  &  -  &  -  &  -  &  -  \\
21 & NGC 2451B & 75.8$\pm$15.2 & $225.2_{-61.3}^{+61.8}$ & -6.9$\pm$ 1.6 & 68.9$\pm$15.3 & 545.7$\pm$202.4 & $0.00_{-0.00}^{+1.00}$ & 0.20 & 0.05 & 0.01 &  -  &  -  &  -  \\
22 & Platais 6 & 61.5$\pm$12.0 & $219.9_{-152.9}^{+156.9}$ & -2.0$\pm$ 3.9 & 59.5$\pm$12.7 & 3806.1$\pm$1302.6 & $6.00_{-6.00}^{+15.00}$ & 0.07 & 0.03 & 0.01 & - & - &  -  \\
23 & ASCC 127 & 66.1$\pm$13.2 & $347.0_{-54.9}^{+54.8}$ & -2.4$\pm$ 8.5 & 63.7$\pm$15.8 & 754.3$\pm$256.4 & $0.00_{-0.00}^{+3.00}$ & 0.43 & 0.29 & 0.08 & - & - &  -  \\
24 & ASCC 100 & 102.4$\pm$20.2 & $348.2_{-58.3}^{+60.0}$ &  0.0$\pm$ 0.0 & 102.3$\pm$20.2 & 431.1$\pm$123.5 & $0.00_{-0.00}^{+0.00}$ & 0.02 & - & - &  -  &  -  &  -  \\
20 & IC 348 & 62.0$\pm$12.4 & $292.1_{-70.9}^{+84.6}$ & -9.6$\pm$ 6.4 & 52.4$\pm$13.9 & 175.2$\pm$53.7 & $0.00_{-0.00}^{+1.00}$ & 0.32 & 0.23 & 0.09 & - &  -  &  - \\ 
22 & Alessi 5 & 51.3$\pm$10.3 & $337.6_{-134.7}^{+101.2}$ & -2.2$\pm$ 3.7 & 49.1$\pm$10.9 & 220.2$\pm$61.8 & $1.00_{-1.00}^{+2.00}$ & 0.13 & 0.05 & 0.02 & - &  -  &  - \\
23 & ASCC 24 &  9.2$\pm$ 1.8 & $249.6_{-70.5}^{+78.8}$ & -8.9$\pm$ 1.8 &  0.3$\pm$ 0.8 & 76.6$\pm$39.3 & $0.00_{-0.00}^{+1.00}$ & 0.26 & 0.26 & 0.21 & 0.01 &  -  &  -  \\
24 & Platais 12 & 169.6$\pm$33.9 & $383.5_{-92.4}^{+64.9}$ & -2.1$\pm$ 5.0 & 167.5$\pm$34.3 & 6506.1$\pm$1304.5 & $0.00_{-0.00}^{+0.00}$ & - &  -  &  -  &  -  &  -  &  -  \\ 
25 & NGC 2516 & 120.3$\pm$24.1 & $310.0_{-68.1}^{+67.2}$ & -6.9$\pm$ 0.8 & 113.4$\pm$24.1 & 1104.3$\pm$261.2 & $0.00_{-0.00}^{+0.00}$ & - & - & - &  -  &  -  &  -  \\
26 & NGC 1746 & 154.9$\pm$30.9 & $367.7_{-199.5}^{+110.8}$ & -12.1$\pm$14.9 & 142.8$\pm$34.4 & 2493.7$\pm$683.1 & $0.00_{-0.00}^{+0.00}$ & - & - & - & - &  -  &  -  \\
27 & Alessi 20 & 165.5$\pm$32.8 & $451.8_{-69.4}^{+71.8}$ &  0.0$\pm$ 0.0 & 165.5$\pm$32.9 & 881.4$\pm$260.9 & $0.00_{-0.00}^{+0.00}$ & - &  -  &  -  &  -  &  -  &  -  \\
28 & NGC 1977 & 12.2$\pm$ 2.4 & $183.4_{-83.9}^{+110.7}$ & -11.7$\pm$ 2.2 &  0.5$\pm$ 1.1 & 75.8$\pm$51.4 & $1.00_{-1.00}^{+1.00}$ & 0.36 & 0.36 & 0.26 & 0.04 & - & - \\
29 & NGC 6405 & 81.4$\pm$16.3 & $487.3_{-96.1}^{+97.0}$ &  0.0$\pm$ 0.0 & 81.4$\pm$16.4 & 537.8$\pm$146.1 & $0.00_{-0.00}^{+0.00}$ & 0.04 & - &  -  &  -  &  -  &  -  \\
30 & NGC 2422 & 131.5$\pm$26.2 & $178.6_{-56.9}^{+71.8}$ & -9.8$\pm$ 1.9 & 121.7$\pm$26.3 & 1168.2$\pm$285.9 & $0.00_{-0.00}^{+0.00}$ & - & - & - &  -  &  -  &  -  \\
31 & ASCC 21 & 12.8$\pm$ 2.6 & $237.8_{-110.8}^{+107.6}$ & -12.6$\pm$ 2.5 &  0.2$\pm$ 0.3 & 1052.5$\pm$427.2 & $5.00_{-3.00}^{+3.00}$ & 0.98 & 0.90 & 0.27 & 0.01 &  -  &  -  \\
 $\vdots$ &  $\vdots$ &  $\vdots$ &  $\vdots$ &  $\vdots$ &  $\vdots$ &  $\vdots$ &  $\vdots$ &  $\vdots$ &  $\vdots$ &  $\vdots$ &  $\vdots$ &  $\vdots$ &  $\vdots$ \\

\hline
    \end{tabular}
    \caption{Sub-list of OCs with overall results of our MC simulation. The full list will be available at the \citep[CDS/vizier-portal;][]{vizier2000}. Columns 1 to 7 shows: the name of the OC, its age, its minimum distance $\pm1\sigma$ percentile, time before present (BP) of minimum distance, age of OC at minimum distance, the clusters birth mass, and the median number of SN produced by the OC $\pm$25\% percentile. Columns 8 to 14 shows the probability that the a SN went off within the OC while the OC was at a distance to the Sun of 1kpc, 0.8kpc, 0.6kpc, 0.4kpc, 0.2kpc, and 0.1kpc respectively. The lower and upper $1\sigma$ percentile we define as the 16\% and 84\% percentile respectively.}
    \label{tab:nearestOC}
\end{table}
\end{landscape}

\subsection{Spatial and temporal supernovae map}
\begin{figure}
	\includegraphics[width=\linewidth]{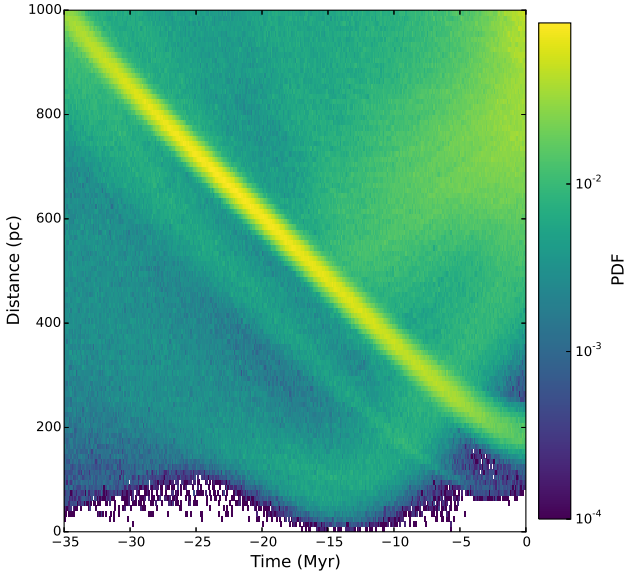}
    \caption{A spatial and temporal probability density distribution of SNe from OCs from -35 Myr to present. Clear structure is seen in the PDF distribution which is the contribution of single OCs. Particular noticeable is Melotte 20}
    \label{fig:SN2dPDF}
\end{figure}

Independent of OC membership, counting all SNe produced within 1 kpc of the Sun yields the 2-dimensional PDF shown in Fig. \ref{fig:SN2dPDF}. Here the binsize is 10\,pc $\times$ 0.1\,Myr. The map shows clear structure which is the trademark of individual OCs, see sec. \ref{sec:individual_ocs}. It also shows that SNe are not homogeneous distributed within the Galactic disc but is most likely closely related to the dynamics of the OC in which they form. Another feature is that at the current time, the probability of a SN increases with distance. Except for the time between -20 and -5 Myr, there is 0 probability for SN within 100 pc during the past 35 Myr. There is very little probability of a SN having exploded within 10 pc at any time during the latests 35 Myr.
% Supernova rate within 1kpc of the Sun per unit time.

\subsection{Past SN rate}
\begin{figure}
	\includegraphics[width=\linewidth]{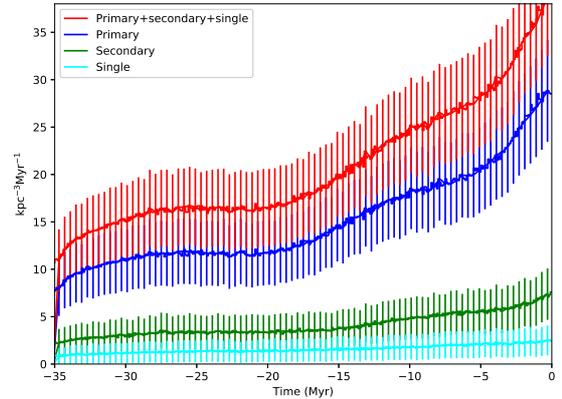}
    \caption{The voluminous SN rate (red curve) within 1 kpc of the Sun in units of kpc$^{-3}$ Myr$^{-1}$ for the past 35 Myr from OC data decomposed into SNe from primary stars (blue), secondary stars (green), and single stars (cyan). The SN rate is computed from counting number of events in bins of 0.1 Myr. The error bars are the Poisson error for each line. To avoid negative numbers we truncate the lower error to yield a SN rate of 0 if necessary.}
    \label{fig:SNrate}
\end{figure}

Binning all SNe within 1\,kpc together in bins of 0.1\,Myr, correcting for the volume (4/3$\upi1\rm kpc^3$) they span and the bin width, yields the SNe rate shown in Fig. \ref{fig:SNrate} in units of kpc$^{-3}$ Myr$^{-1}$. The total SN rate is shown in red. Its components are SNe from primary stars (blue), SNe from secondary SNe stars  (green), and SNe from single stars (cyan). Binary primary star SNe are by far the most dominating component with a relative frequency of $74\%$, while binary secondary star SNe the second most dominant component, yielding $20\%$ percent contribution. It is worth noticing that single star SNe are few and indeed rare with only $6\%$. That single star SNe contribute so little is because the binary fraction for SN progentiros is nearly unity. The associated error is from Poisson statistics.
The current rate averaged over the past 1 Myr is $37.8\pm6.1$ SN kpc$^{-3}$ Myr$^{-1}$. Going back in time the reconstructed SN rate drops due to the decay of OC populations as they age. This decay introduces an observational bias in the sense that the older a population of OCs we consider the fewer have survived to be observed to day. This directly affects the number of SNe that can be estimated directly at some point back in time. 

% Supernova rate within 1kpc of the Sun per unit time.

\subsection{Interesting open clusters}\label{sec:individual_ocs}
In this section we highlight some of the most interesting OCs based on their estimated contribution of SNe. The results we present is based on data from table \ref{tab:data} and relevant literature.

\subsubsection{Mamajek 1 or the $\eta$ Chamaeleontis cluster}
\citet{mamajek1999} reported on the discovery of a compact young, <18\,Myr, OC close to the Sun at 97\,pc. Its most prominent member is the B8 star $\eta$ Chamaeleontis. Being the first cluster discovered via X-ray observations, Mamajek 1 is a peculiar cluster. Its stellar content consist of several weak T Tauri stars with an age spread between 2-18\,Myr \citep{mamajek1999} and it is deficit in wide binaries \citep{brandeker2006}. Further studies concludes that Mamajek 1 most probable had its origin in the Scorpious-Centaurus OB association some 6.7 Myr ago \citep{mamajek2000,jilinski2005}.
Based on extrapolation of initial stellar mass functions and known members above 1\msun{} \citet{mamajek2000} estimates a current total mass of $31 \pm 14$\msun.
On comparison with the data applied from table \ref{tab:data} it is clear that Mamajek 1 is more massive than we have applied.

\subsubsection{Melotte 20 or The $\alpha$ Persei Cluster}
The Melotte 20 or $\alpha$ Persei cluster is the most prominent OC in our simulations owing to its accurate space velocity measurements and apparent high mass. The $\alpha$ Persei cluster had a birth mass of $\sim$9100\msun{} and contributed with $\sim$87 SNe in the past 35 Myr. We find that most probably these SNe went off at distances greater than 500 pc. The current mass from \citet{Piskunov:2008ft} which we used is 7900\msun. A literature search of the $\alpha$ Persei cluster suggests a current mass of 352\msun, which is much less, and an age factor of 2 older \citep{sheikhi2016}. The age estimate of the cluster is however not well constrained \citet[see for instance][]{zuckerman2012,silaj2014}. This obviously conflicts with our adopted initial conditions and obtained results for this cluster. Given its high contribution in SNe, it might also introduce biases in the overall estimate of the SN rate back in time. Member detection and hence mass estimate of the $\alpha$ Persei cluster is particularly difficult because its low galactic latitude projects it onto a high density stellar field and separation among different stellar groups is thus challenging \citep{prosser1992}.
By not taking the OC and its SN production into account only has an effect for the SN history at earliest times, i.e. prior to 20 Myr BP.

\subsubsection{ASCC 19}
ASCC 19 was identified in a search of OCs within the Tycho catalogue by \citet{Kharchenko:2005ba, Kharchenko:2005dw}. Since then we found no case study of this OC except for those studies applying OC catalogue data to study the OC population.
From our simulations we find the minimum distance to have been $51_{-32.4}^{+81.0}$pc at $17.8\pm8.6$ Myr BP. Its birth mass is estimated to be $M_{\rm birth} = 721\pm256$\msun{} and it has produced 3 to 8 SNe since birth. Interestingly we find a 61\% probability that a SN was within 400 pc and 17\% probability that the SNe was within 200 pc of the Sun. There is only a 1\% chance that one of its SNe was within a distance of 100 pc.

\subsubsection{NGC 2232}
\citet{silaj2014} determined the age of NGC 2232 to $32\pm4$Myr using Geneva isochrones which fit well with our applied age. Their best fit is obtained by modulating the magnitude of the stars as if the OC is closer than estimated with parallaxes fom the reduced Hippachos \citep{vanleeuwen2009}. But \citet{silaj2014} do not give a new distance.
We find that NGC 2232 was closest to the Sun $12.8\pm7.8$ Myr BP at a distance $D=103.7_{-33.9}^{+189.8}$pc, and with a birth mass of $M_{\rm birth} = 245\pm85$\msun{}, has contributed 2-4 SNe since its birth. We find a 63\% probability that a SN occured within 400 pc, 18\% probability that it was within 200 pc and 2\% probability that a SN was within 100 pc.

\subsubsection{Roslund 5}
Roslund 5 is a little studied OC despite being discovered already in 1960 \citep{roslund1960}. The fact that it has not been more carefully studied is probably due to a later photometric study by \citet{leeperry1971}. \citet{leeperry1971} measured a large scatter in the color magnitude diagram and concluded that probably, Roslund 5 was not an OC. Using Hipparchos observations the nature of Roslund 5 was revisited by \citet{baumgardt1998} who by including proper motions and parallax distances concluded that Roslund 5 probably is an OC. Later OC identifications pipelines also identify OC-like properties in the position of Roslund 5 which only seems to confirm that indeed Roslund 5 is an OC \citep{Dias:2002fq,Kharchenko:2005ba}.
We find Roslund 5 to currently being at its closest distance to the Sun since birth. It was formed with an initial mass of $M_{\rm birth} = 567 \pm 164$\msun{} and has contributed with 0 to 3 SNe since birth, however all beyond a distance of 400 pc.

\subsubsection{NGC 1981}
The OC NGC 1981 is lying in the foreground of the Orion nebula and is believed to be a binary OC companion gravitational bound to NGC 1976 \citep{alves2012,priyatikanto2016}. A case study puts its mass to be $137\pm14$\msun{} and an age of only $5\pm1$Myr which disagress with the age used here of $\approx$30 Myr. The present mass we have used is much less than that given by \citet{alves2012} and we estimate a birth mass $M_{\rm birth} = 114\pm48$\msun{} and a likely SN production of 1-2 SN. NGC 1981 is very interesting as we find a probability of 30\% of it having set its SNe within 200pc and 9\% probability that it was within 100 pc. The suggested time of closest approach of $66.3_{-31.6}^{+36.6}$pc was some 12.8$\pm$3.1 Myr BP.

\subsubsection{NGC 1976 or The Trapezium Cluster}
NGC 1976 also known as the Trapezium Clusters, located in the Orion cloud, is an active star forming region and has been for several Myr. Being one of the most massive star forming regions within its distance to the Sun and having very little extinction makes this region a prime candidate for any observational program on (low mass) stellar formation \citep{kubiak2017}. We find from our simulations of the OC's motion back in time, that is has been closer to the Sun, likely within 100 pc, with a 1\% chance of setting of a SN during this time. Overall we estimate the cluster to have produced 1-3 SN during the past 35 Myr. Relative to our ages from \citet{Piskunov:2008ft} of $\sim$50 Myr, the age given by \citet{kubiak2017} is a factor of $\sim$5 smaller. We estimate the birth mass of the OC to be 352$\pm$134\msun. The minimum distance we find is $57.3_{-29.2}^{+46.0}$ pc at $12.8\pm2.9$ Myr BP. A cause for precaution is the large spread in age estimates found in the literature. If the region is undergoing sequential star formation, its star-formation activity can reach far back in time and this could explain the discrepancy in different age estimates. 

\subsubsection{Trumpler 10}
Despite being part of many studies, we do not find a study dedicated to the stellar content of Trumpler 10 which is actually much needed as this OC is very interesting with respect to recent SN. \citet{schilbach2008} find for instance that $\zeta$ Puppis might have been ejected from this OC some 2.5 Myr ago and \citet{tetzlaff2010} find evidence that the NS pulsar RX J0720.4$-$3125 was born either in TW Hydrae 0.4 Myr ago or Trumpler 10 0.5 Myr ago. 
We find Trumpler 10 to have had a birth mass of $M_{Birth} = 722\pm259$\msun{} and produced 4 to 8 SNe which all were beyond 200 pc.

\subsubsection{ASCC 20}
ASCC 20, like ASCC 19 is not studied in the literature except for those studies of the Galactic OC population. Our simulations suggests the OC had its closets passage to the Sun $13.1\pm7.1$ Myr BP though its minimum distance, $D = 101.1_{-50.1}^{+284.2}$pc, is not well constrained. Its estimated birth mass $M_{\rm birth} = 954\pm369$ \msun{} suggests that it can have produced somewhere between 5 and 10 SNe all beyond 200 pc, except for a 3\% probability that a SNe was between 200 and 100 pc.

\section{Discussions}
Here we discuss the uncertainties and the potential impact of our results. The impact is discussed in relation to the known literature on Galactic SN frequencies, the solar neighbourhood SN rate, near Earth supernovae, and predicted potential impacts on Earths climate and geology. Uncertainties are discussed with respect to the used data and the physics omitted.

\subsection{Mass estimate of open clusters}
Potentially, the estimated masses of OCs from \citet{Piskunov:2008ft} are biased due to projection effects or simplifications in the method applied \citep{Ernst:2010hg}. Alternative data sources using for instance the number of cluster members might suffer from other biases that has the same overall effects or are incomplete. For example, in the recent catalogue of \citet{Dias:2014fr} the OC Mamajek 3, only 90 pc from the Sun, has some ~20000 members, an absolute proper motion of ~0\kms and a radial velocity of 18.5\kms. These numbers would suggest this particular OC to have come within 10 pc of the Sun around 5 Myr BP and over the cause of its 20 Myr life time have deposited several tens of SNe very close to the Sun. 
A color magnitude diagram of the stellar population indicates that Mamajek 3 in the \citet{Dias:2014fr} catalogue is conflating at least two stellar populations. 
Estimating to completeness the stellar content of an OC is in general very difficult and we are so far forced to accept high uncertainty in member populations and hence also mass estimates. 
Our approach of applying MC ensembles is not only necessary to account for the problem but it also allows for a proper statistical account of the uncertainties within the data which are fairly large. 
Systematic errors in data processing effects is however beyond what can be accounted for using MC simulations and also beyond the scope of the present paper. Finally, we highlight that the result for Melotte 20 is probably due to systematic biases in the method of \citet{Piskunov:2007bg} as pointed out by \citet{Ernst:2010hg}.

\subsection{Position of SN progenitors within OC}
We have implicit assumed that the position of the SN is given by the position of the center of the OC at the time of the SN. Hereby we ignore the spatial stellar distribution an OC has which introduces some systematic uncertainty in the actual position of the SN. However, the uncertainties in the position and velocity of each OC in our sampling indirectly mimics this spatial stellar distribution why using the simple interpolation scheme onto the trajectory is a robust solution to estimate the position of the SN.

\subsection{The distribution of primaries, secondaries and single star supernovae}
Figures \ref{fig:delay_time} and \ref{fig:SNrate} show the decomposed number and rate of SNe. The number of single star SN progenitors are few relative to those born in binaries and is a consequence of initial stellar mass priors used in relation to the population synthesis modelling described in sect. \ref{sec:priors} and \ref{sec:PopSynth}. If star formation actually produces SN progenitors as single stars or if these become singles due to dynamical ejection scenario or binary supernova ejection is not known. The percentage of single star SN progenitors is $\sim$6\% in our simulation. This is roughly equal to the  fraction of runaway SN progenitor stars to bound SN progenitor stars \citep{eldridge2011}.
The number of secondary stars, i.e. the companion of a binary that goes SNe is $\sim.$15\% which is a factor 2.5 less than estimate in \citet{eldridge2011} but is in agreement with that of \citet{zapartas2017}. Likely the main reason for the reduced fraction of SN progenitors from secondary stars is the differences in priors from on the one side \citet{eldridge2011} and \citet{zapartas2017} and this study on the other side.

\subsection{Effects of runaway stars}
Some of the SN progenitors formed in embedded clusters will be ejected from their birth environment either due to dynamical interactions or by being the initially less massive star in a binary that is disrupted due to the initial more massive star going SN \citep{blaauw1961, hoogerwerf2001}. Runaway OB-type stars can potentially introduce a spread in the distribution of SNe that goes beyond the uncertainties in the OC data.
However, these stars are only a small fraction of all SN progenitors. \citet{eldridge2011} estimates that only about 0.5-2.2\% of all O-type stars within the Milky Way will be runaway stars originating from a binary ejection scenario while they point out that 4\% of O-type stars in the catalogue of \citet{maiz2004} are identified as runaway stars. \citet{tetzlaff2011} looked at young (<50Myr) stars, and identify a runaway frequency among these young stars of 27\%, but they do not give the fraction of runaway stars among SN progenitors.

%\begin{equation}
%N = \sum S(t) \sum \psi_i \sum\delta_i(t-\tau)\Delta \tau
%\end{equation}
%N is number of stars, S(t) is the star formation rate, $i$ is number of systems, $\psi$ is the population of stars, the delta function selects systems in the phase of interest, i.e. SN.
\subsection{The observed decay in the estimated SN rate}
\begin{figure}
    \begin{tabular}{l}
        \includegraphics[width=\linewidth]{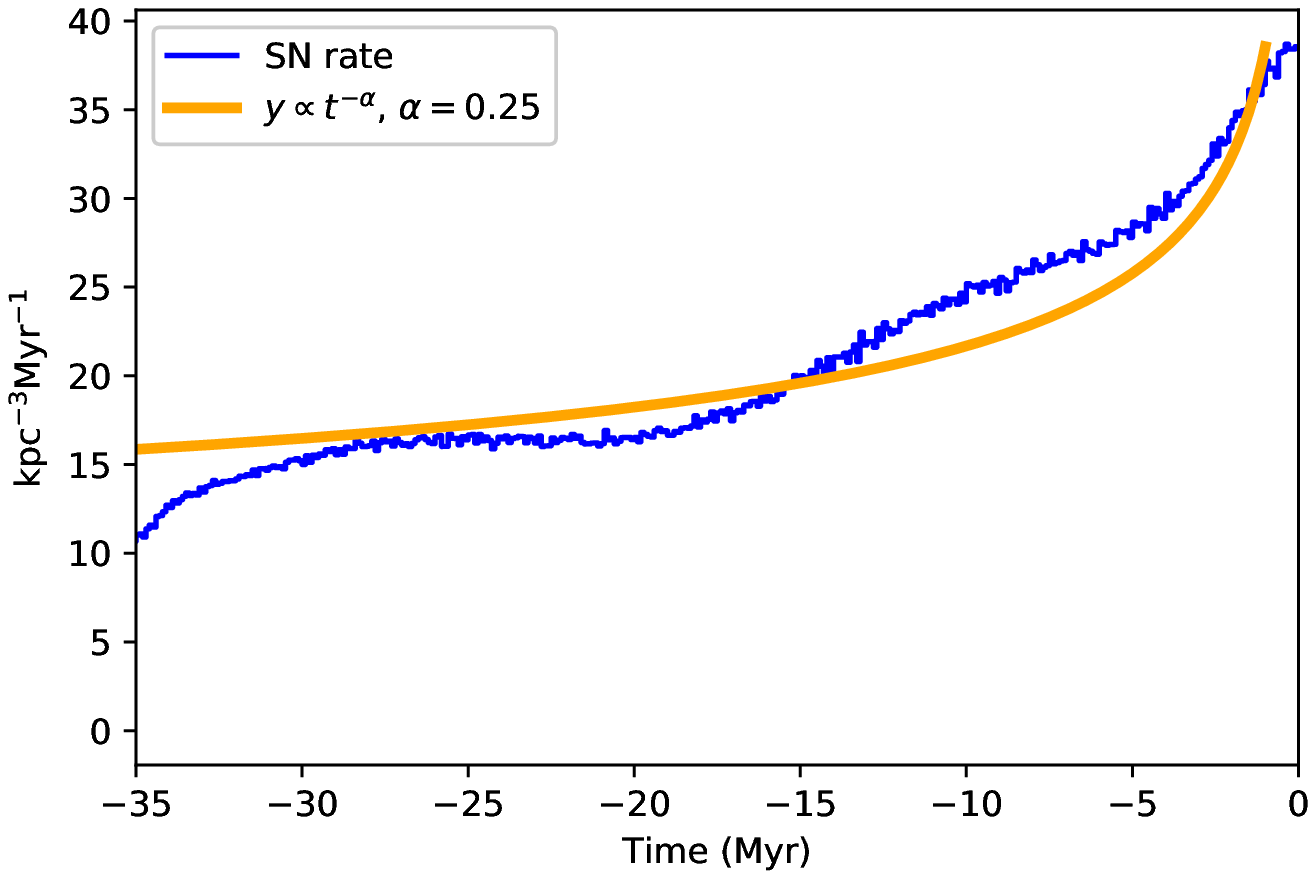}     \\
        \includegraphics[width=\linewidth]{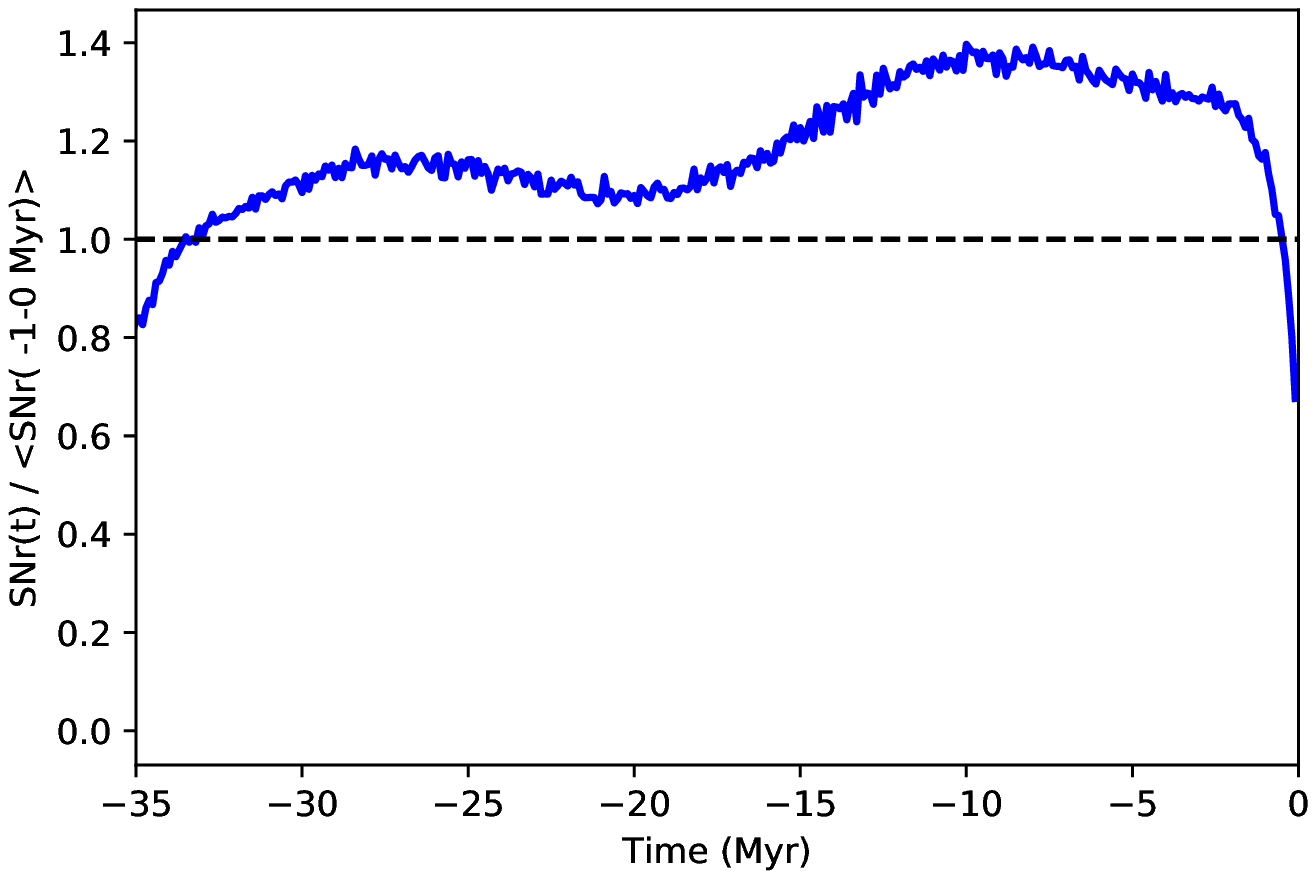}
    \end{tabular}
    \caption{Top panel is the same as the total SN rate shown in Fig. \ref{fig:SNrate} with a power law function $y \propto t^{-\alpha}$ with $\alpha$=0.25 plotted on top. The lower panel shows the relative SN rate back in time relative to the most recent 1 Myr by convolving the decay function with the reconstructed SN rate.}
    \label{fig:SNrate_relative}
\end{figure}

The SN rate shown in Fig. \ref{fig:SNrate} is seen to decrease as one goes back in time. This is an observational bias as only a fraction of the initial population born some time in the past is still observable. The remaining fraction is dissolved. We can translate the decay of an OC generation into the decay in SNe events that can be estimated from the same population of OCs. In this respect, older generations of OCs contribute with fewer and fewer of the SNe events estimated in figs \ref{fig:SN2dPDF} and \ref{fig:SNrate}.
If we assume that the SNe that can be estimated within an OC generation of some age decays following a power law like that of OCs \citep{delaFuenteMarcos2008} for a constant star formation rate, then it is valid to assume that the fluctuations around this decay curve is local fluctuations in the star formation history in the past. Following the method of \citet{svensmark12} we compute the relative SN rate in past as shown in Fig. \ref{fig:SNrate_relative}.
% Supernova rate within 1kpc of the Sun per unit time.
Relative to today the SN rate in the past show only small fluctuations though a clear shift in around 20 Myr BP is observed.

\subsection{On the deposition of \fesixty{} from supernovae originating in the Scorpius-Centaurus OB association}
Figure \ref{fig:SN2dPDF} shows that no SNe is expected within 100\,pc of the Sun during the past 5 Myr. At first sight this is in contrast to the observation of \fesixty{} anomalies from lunar- and terrestrial sources \citep{knie99,knie04,Fitoussi:2008cb,wallner16}. There are at least two explanations for this. First, the OC or OB association that may have produced the close SNe roughly 2.8-2.2 Myr ago is not found within our data set given in Tab. \ref{tab:data}. Secondly, the general assumption that the SNe happened inside a known OC or association need not be true. Alternative sources could be be a high velocity runaway star, although unlikely, further as is elaborated below, binary evolution might play a role as well.
The suggestion put forth that the Scorpius Centaurus (Sco Cen) OB association \citep{benites02,breitschwerdt16} is the probable parent of the most recent near-Earth SNe do not account for effects of binary evolution or dynamical ejection of SN progenitors. But such effects might be needed. In particular, the proposal that two  SNe from single star evolution, of 9.2$M_{\rm \odot}$ and 8.8$M_{\rm \odot}$ respectively, happened within just 0.8 Myr apart from each other, within Sco-Cen is hard to reconcile with the young age of Sco-Cen which is estimated to be at most 16 Myr \citep{preibisch2008}. A SN progenitor below 10\msun{} lives a considerable amount of years beyond 16 Myr. However, if the yield of \fesixty{} from a 9.2$M_{\rm \odot}$ and a 8.8$M_{\rm \odot}$ SN progenitor 0.8Myr apart reproduces the observed data, then it is necessary to invoke a stellar evolution scenario that allows a correct yield and is consistent with the age of the Sco-Cen. One could imagine that binary evolution, in the form of increased mass loss due to Roche lobe overflow from a SN progenitor in a binary could consistently explain the \fesixty{} yields and the age of the Sco-Cen simultanously.
Finally, if the distance to the SNe needed in order to properly reproduce the \fesixty{} signal is not fitting with the most probable distance of the SN progenitor(s) from a single star evolution perspective \citep{fry2015}, dynamical ejections and binary evolution could offer a broader range of potential solution of which some may provide a more satisfying solution. With the high fraction of binary among SN progenitors, as seen in fig. \ref{fig:priors}, it seems very realistic that binary evolution can have an important role in explaining the \fesixty{} signal.

\subsection{Supernova rates from $\rm ^{14}$C and $\rm ^{10}$Be}
Using $\rm ^{14}C$ anomalies in the past 50 kyr and the $\rm ^{10}Be/^{9}Be$ isotope anomaly time series going back to 300 kyr BP \citet{Firestone:2014cj} suggested that 23 SNe within 300 pc went off in the past 300 kyr, see also \citet{melotte2015}. Our simulations do not reproduce this. The voluminous SN rate required to yield the high frequency can be computed as 23/((4/3)$\upi \times$0.3 kpc$^3$ $\times$ 0.300 Myr) = 677 kpc$^{-3}$ Myr$^{-1}$ which is far from our computed SN rate during this time.
It has been previously suggested that the SN rate in the solar neighbourhood is high \citep{Dragicevich:1999ep,Grenier:2000vw} however, it is not in agreement with our observations of MW type galaxies \citep{Tammann:1994ir} and the \gr{}-background of $\rm ^{26}Al$ \citep{Diehl:2006fr} unless we are in a privileged position in the MW with respect to SNe. Turning it around, since, \citet{Firestone:2014cj} uses cosmic-ray spallation production in the atmosphere as a proxy for SNe, it is worth asking if another closer source such as the Sun could produce a similar signal \citep{svensmark2000}.

\section{Conclusions}
We have compiled a data set of 395 OCs for which we have reconstructed their birth mass, single- and binary stellar evolution and tracked their distance to Sun for the past 35 Myr. We find that several OCs have come close to the Sun while having an age allowing them to have SN progenitors.
In particular we identify ASCC 19, NGC 1981, NGC 1976 to very likely have deposited a SN within 200 pc of the Sun and with somewhat smaller probability within 100 pc in the period between 17 and 12 Myr BP.
From our simulations we also produce a spatio-temporal 2D PDF map showing that the solar neighbourhood distribution of SNe has structure and is not uniformly distributed. In this perspective it is likely that the distribution of near-Earth SNe are correlated with their birth cluster. However, as we discuss, effects such as dynamical and binary ejection of SN progenitors from OCs will introduce a small spread in the SN distribution which we do not account for.
The detected $\rm ^{60}Fe$ signal from recent near-Earth SNe events $\sim$2.8-2.2Myr BP, we suggest, could be better explained if one considered binary evolution. Hereby the estimated age of Sco-Cen would not be violated relative to using single star evolution. Indeed this would also better fit with most SN progenitors being born in binaries.
Despite differences between the applied data and case studies of OCs, the results seems robust. It is worth noticing that these differences come from comparing OC identification pipe lines which will never be as detailed as studies dedicated to individual OCs. On the other hand, they provide data obtained in a consistent manner across the applied population.
With the release(s) of data from the \citep[GAIA;][]{gaia2016} satellite and bi-products hereof, new identifications of OC membership, improved parallax, and proper motion will allow for a greater membership completeness which in turn will constrain the mass of solar neighbourhood OC population. Hence better constrain the solar neighbourhood SN history and possibly extend the historical time that can be estimated. Finally, GAIA will likely lead to the discovery of hitherto unknown OCs, hereby extending the distance out to which our sample of OCs are complete.

\section*{Acknowledgements}
MS thanks Maxwell Moe for sharing his IDL script to sample the binary population from the priors given in \citet{moe2016} and acknowledges support from the Swiss National Science Foundation project ID: 200020\_172505. Finally we wish to thank Tassos Fragos, Georges Meynet, Thomas D\o ssing, Jan Faye, Andrii Neronov, Christoffer Karoff, Nami Mowlavi, and Livia Kother for fruitful discussions and support.

%\begin{appendix}
%\section{Past SN with revised mass of the Melotte 20 / the $\alpha$ Persei cluster}\label{sec:appendix}
%\end{appendix}

%%%%%%%%%%%%%%%%%%%%%%%%%%%%%%%%%%%%%%%%%%%%%%%%%%


\bibliographystyle{mnras}
\bibliography{library} % if your bibtex file is called example.bib

%%%%%%%%%%%%%%%%%%%%%%%%%%%%%%%%%%%%%%%%%%%%%%%%%%

% Don't change these lines
\bsp	% typesetting comment
\label{lastpage}
\end{document}